\documentclass[12pt]{article}
\usepackage{amsmath}

\def\ex#1{\langle#1\rangle}
\def\W#1{W_{\text{#1}}}
\def\tr{\mathop{\text{tr}}}

\def\SU{SU}

\def\Nis{\mathcal{N}=}

\begin{document}
\thispagestyle{empty}
\hbox{}

\begin{flushright}
UT-02-60\\
hep-th/0211189
\end{flushright}

\bigskip\bigskip\bigskip
\bigskip\bigskip\bigskip

\begin{center}
\LARGE

Derivation of the Konishi anomaly relation
from Dijkgraaf-Vafa with (Bi-)fundamental matters

\bigskip

{\Large
Yuji Tachikawa}\\

\normalsize
\bigskip

\textit{
Department of Physics, Faculty of Science, University of Tokyo,\\
Hongo 7-3-1, Bunkyo-ku, Tokyo 113-0033, Japan}

\bigskip

email: \texttt{yujitach@hep-th.phys.s.u-tokyo.ac.jp}
\end{center}

\textbf{Abstract:}
We explicitly write down the Feynman rules
following the work of Dijkgraaf, Vafa and collaborators
for $\Nis1$ super Yang-Mills
having products of $\SU$ groups as the gauge group
and matter chiral superfields in 
adjoint, fundamental, and bi-fundamental representations
without baryonic perturbations.
As an application of this,
we show expectation values
calculated by these methods
satisfy the Konishi anomaly relation.

\newpage\setcounter{page}{1}
\section{Introduction}
Dijkgraaf and Vafa \cite{Dijkgraaf:2002dh} proposed a duality
between four dimensional $\Nis1$ super Yang-Mills theories
and the old matrix model, which states that the effective
superpotential of the super Yang-Mills written as a function of
the gaugino condensate $S$ and coupling constants,
can be obtained from a matrix model free energy.
This important proposal is checked by 
many papers \cite{CHECKS}.

Originally the proposal were made in the context of 
the deformation of $\Nis 2$ theories,
recent papers \cite{Flavors}
showed that it can be generalized to include 
models with fundamental matter, 
and successfully reproduced the
Affleck-Dine-Seiberg superpotential by this method.
But it seems to us there is some confusion about what is the correct
prescription.
This can be determined if one follows the argument
in the paper by Dijkgraaf, Grisaru, Lam, Vafa and Zanon (DGLVZ) \cite{Dijkgraaf:2002xd},
in which they derive
the Dijkgraaf-Vafa proposal by
integrating out matter superfields in the presence of
an external gauge field.
Although the analysis made in DGLVZ is quite clear for
diagrams containing at least one vertex, it needs some care
in evaluating the contribution from one loop diagram without any
vertex.

The purpose of this short note is
to write down the precise rules for calculating the 
effective superpotential and expectation values for models 
having products of $\SU(N_i)$ as the gauge groups and matters
in the fundamental, adjoint,
or bi-fundamental representation and without baryonic perturbations.
The restriction to the case without baryonic perturbations
comes from the fact that
they cannot be depicted by 't~Hooft's double-line notation.
For the case with only quadratic mass perturbations,
this method reduces to that presented in
the section three in Intriligator's `integrating in' paper
\cite{Intriligator:1994uk}.

As an application we show the expectation values
calculated by the prescription of this paper
satisfies important identities coming
from the Konishi anomaly\cite{Konishi:1983hf}.
The importance of the Konishi anomaly
in the framework of Dijkgraaf-Vafa
was first pointed out by Gorsky\cite{Gorsky:2002uk}.

In the following, we follow the conventions of \cite{Dijkgraaf:2002dh}.

Note Added:
\quad
After completion of this work, we have noticed a new preprint
hep-th/0211170 by Cachazo, Douglas, Seiberg and Witten \cite{CDSW},
which has some overlap with this article. 

\section{Prescription}
First, recall that DGLVZ explicitly states the following Feynman rules for
the contribution of a diagram to
the effective superpotential of the gaugino condensate: \begin{enumerate}
\item Write the diagram following 't~Hooft's double-line notation.
\item Assign on each index loop the gaugino condensate 
$S$ or the dimension of the fundamental representation $N$.
\item The contribution to the superpotential comes only from
those diagrams, in which the number of $S$ assigned
is equal to the number of independent loop momenta.
\item The propagator for each matter superfield is the inverse of its mass.
\item Interaction vertices come from the cubic and higher order 
terms in the tree-level superpotential.
\item Finally, multiply them together.
\end{enumerate}

The restriction 3 ensures
the correct number of Grassmann integrals,
because there are two Grassmann integrations for each momentum loop
and each insertion of $S$ contains two Grassmann variables.
The same condition restricts the topology of the diagrams
that can contribute to the effective superpotential
to be a sphere or a disk.

Let $\W{eff}$ denote the 
non-perturbative
superpotential calculated following Dijkgraaf-Vafa.

The expectation value of the lowest component of a gauge-invariant
operator can be calculated using the above rules and finally substituting
$S$ by the value which extremizes $\W{eff}$.

Now let us determine the contribution of one-loop graphs with no insertion.
\begin{itemize}
\item For $Q$ in the fundamental and $\tilde Q$ in the antifundamental
of $\SU(N)$ with a mass term $mQ\tilde Q$, the expectation value of
the bilinear satisfies
$\ex{Q\tilde Q}=S/m$. This is equal to $m(\partial/\partial m)\W{eff}$,
so that $\W{eff}$ contains $S\log (m/\Lambda)$, where
the integration constant $\Lambda$ is some multiple of 
the 
dynamically generated scale
for pure $\Nis1$ super Yang-Mills.
We adopt a prescription that the proportionality constant is unity.
\item The case with $\Phi$ in the adjoint. The argument goes essentially the
same with that given above,
except that the diagrams are now double-lined and so
the 
loop contributes \[
NS\log (m/\Lambda).
\]
\item A bi-fundamental $\Phi$ for $\SU(N_1)\times \SU(N_2)$. Its effect for 
$SU(N_1)$ 
is the same as introducing $N_2$ fundamentals of $\SU(N_1)$.
\end{itemize}

And finally, as noted by Dijkgraaf and Vafa, we must add
the Veneziano-Yankielowicz term\cite{Veneziano:1982ah}
 $NS(1-\log(S/\Lambda^3))$
for each gauge group.
The total effective superpotential can be written as \[
\W{eff}=\W{VY}+\W{one loop}+\W{higher},
\]where $\W{VY}$ is the Veneziano-Yankielowicz piece,
$\W{one loop}$ is what are discussed in the preceding paragraph,
and $\W{higher}$ comes from diagrams containing at least one vertex.
With only quadratic perturbations, this formula
reduces to that mentioned in \cite{Intriligator:1994uk}
section 3.

As a consistency check,
one can 
verify  that the prescription given here correctly reproduces
for example
the decoupling equation for fundamental flavors\[
\Lambda_{\text{after}}^{3N-N_f+1}=
m\Lambda^{3N-N_f}_{\text{before}}.\]

\section{An Example}
As an example, consider the $\Nis1$ supersymmetric 
$\SU(2)_1\times \SU(2)_2$ theory with 
two fundamentals $L_1, L_2$ of the second $\SU(2)$ and 
one bi-fundamental $Q$.
(This model is the one given in \cite{Intriligator:1994jr} section 4.1 and
the superpotential (\ref{X}) as a function of
$S$ and $m$ is 
essentially given in \cite{Intriligator:1994uk} section 3,
so this is not essentially new. It is just for an illustrative purpose.)
Let us denote the gaugino condensates and the dynamically generated scales
of each group
as $S_1$, $S_2$ and $\Lambda_1$, $\Lambda_2$, respectively.
Then include mass perturbations $mQ^2+\mu L_1L_2$ to the tree level
superpotential.
As the bifundamental is $(\hbox{\boldmath$2$},\hbox{\boldmath$2$})$,
we can write down immediately the effective superpotential of
the gaugino condensate as
\begin{equation}
\W{eff}=2S_1(1-\log\frac{S_1}{\Lambda_1^3})
+2S_2(1-\log\frac{S_2}{\Lambda_2^3})
+S_1\log\frac{m}{\Lambda_1}
+S_2\log\frac{m}{\Lambda_2}
+S_2\log\frac{\mu}{\Lambda_2},
\label{X}
\end{equation} where the first and the second terms 
are the Veneziano-Yankielowicz superpotential for each of the gauge groups,
and the third and the fourth are the contribution of the bi-fundamental,
and the last comes from two fundamentals.
The expectation value of $X=\ex{Q^2}$ and $Y=\ex{L_1L_2}$ 
can also be calculated to be 
$\ex{Q^2}=(S_1+S_2)/m$ and $\ex{L_1L_2}=S_2/\mu$. 
Integrating out $S_i$'s and 
writing $\W{eff}=\W{non-perturbative}+m\ex{Q^2}+\mu\ex{L_1L_2}$,
we immediately obtain\[
\W{non-perturbative}=\ex{S_2}
=\frac{\Lambda_1^5 Y}{{XY-\Lambda_2^4}}.
\]

\section{Derivation of the Konishi relation.}
As remarked by a recent paper by Gorsky\cite{Gorsky:2002uk},
the Konishi anomaly relation,
when evaluated at some supersymmetric vacuum,
amounts  in the Old Matrix model language
to the Virasoro $L_0$ condition supplanted by
an anomaly of entropy terms.
He also checked the relation with several examples
already appeared on the literature as the test for the 
Dijkgraaf-Vafa proposal.

Here we derive, as an easy application of the rules
stated in the previous section,
the Konishi relation from the Dijkgraaf-Vafa prescription.
What we want to prove is that \[
2N\ex{S}=\ex{\phi\frac{\partial}{\partial \phi}W_{\text{tree}}}
\]for each adjoint chiral superfield $\phi$ and\[
\ex{S}=\ex{Q\frac{\partial}{\partial Q}W_{\text{tree}}}
\]for each fundamental $Q$.
We present a derivation for the case of an $\SU(N)$ adjoint.
The proof for other cases 
is essentially the same.

Decompose $\W{tree}$ and $\W{eff}$ as \[
\W{tree}=\frac12m\tr\phi^2+(\text{other mass terms})+\W{vertices},
\] where $\W{vertices}$ is the interaction terms in the 
tree-level superpotential, and\[
\W{eff}=\W{VY}+\W{one loop}+\W{higher}
\] where \begin{align*}
\W{VY}&=NS(1-\log (S/\Lambda^3))+(\text{V-Y terms for other gauge groups}),\\
\W{one loop}&=NS\log (m/\Lambda)+(\text{one loop terms for other flavors}),
\end{align*}
and $\W{higher}$ comes from the diagrams which contains at least one
interaction vertex.

Now, note that
$\phi(\partial/\partial\phi)$ counts the number of 
the $\phi$ legs of each diagram in $\W{tree}$, 
\[
\ex{\phi\frac{\partial}{\partial\phi}\W{vertices}}=
\sum_{\text{diagram}\ D}(\text{the number of legs of $\phi$ in $D$})
\times(\text{the value of $D$}),
\]But the number of $\phi$ legs is twice the number of the $\phi$
propagators, so that \[
\ex{\phi\frac{\partial}{\partial\phi}\W{vertices}}=
2m^{-1}\frac{\partial}{\partial m^{-1}}\W{higher}.
\] On the other hand, the definition of $\W{eff}$ shows \[
\ex{m\tr\phi^2}=2m\frac{\partial}{\partial m}\W{eff}.
\] Here all the partial derivatives on the RHS are taken first,
and then $S$ is substituted by the value which minimizes the $\W{eff}$,
so that\begin{align*}
\ex{\phi\frac{\partial}{\partial \phi}W_{\text{tree}}}
&=2m\frac{\partial}{\partial m}(\W{eff}-\W{higher})\\
&=2m\frac{\partial}{\partial m}(\W{VY}+\W{one loop})\\
&=2N\ex{S}.
\end{align*}This is what we want to derive.

\section{Conclusion and Outlook}
In this short note we explicitly write down the perturbative rules
of the Dijkgraaf-Vafa proposal for classical gauge groups with 
fundamental, adjoint, or bifundamental matter fields,
and showed that the Konishi anomaly relation is indeed
satisfied by the expectation values calculated from these rules.

One of the biggest remaining problems is the incorporation of 
baryonic perturbations which involve the invariant tensor
$\epsilon_{ijk\cdots}$. And another is the extension of this framework
to the chiral matter contents. In this case, propagators cannot be
easily given because no gauge-invariant mass deformations can be introduced.
Really interesting, and also phenomenologically important examples 
of $\Nis1$ super Yang-Mills theory are usually of this kind. So
this well deserves a study.

\paragraph{Acknowledgments}
The author would like to thank T. Eguchi, K.-I. Izawa,
Y. Nakayama and K. Sakai for  very helpful discussions.

\end{document}